\documentclass[11pt]{article}
\usepackage[margin=1in]{geometry}
\usepackage{amsmath,amssymb}
\usepackage{graphicx}
\usepackage{booktabs}
\usepackage{hyperref}

\title{Representation Homogeneity and Systemic Instability in AI-Dominated Financial Markets: A Structural Approach}
\author{Yimeng Qiu, Qiwei Han}
\date{\textit{Draft Version}}

\begin{document}
\maketitle

% ============================================================
% ABSTRACT — Fixed: "hold fixed" → "conditioning on alternative 
% distributions"
% ============================================================

\begin{abstract}
This paper investigates how similarity in the informational representation of market states among Artificial Intelligence (AI) trading agents can generate systemic instability in financial markets. We construct a structural multi-agent market model calibrated using high-frequency microstructural moments. AI agents are modeled through a two-layer decision architecture consisting of a nonlinear representation layer and an adaptive linear readout layer. The representation layer maps raw market states into high-dimensional feature vectors, while the readout layer generates return forecasts that feed into a risk-controlled trading rule. This representation-based microfoundation separates two objects that are often conflated in the literature: \emph{representation homogeneity}---the degree to which agents encode market states into similar feature spaces---and \emph{forecast overlap}---the degree to which agents produce similar return predictions. We show theoretically that these two concepts are related but not equivalent, and that representation homogeneity can compress the effective space of forecast disagreement under stress even when predictions appear diverse in normal times. Through controlled factorial experiments that vary representation homogeneity while conditioning on alternative risk-aversion and learning-rate distributions, we hypothesize that increasing representation similarity amplifies synchronization in beliefs and positions, leading to volatility clustering, liquidity stress, and elevated tail risk. Our structural mechanisms suggest that low perceived volatility regimes can endogenously accumulate hidden leverage through position stickiness, which subsequently collapses when shocks trigger synchronized deleveraging. The results provide a structural foundation for macroprudential policies aimed at monitoring and preserving diversity in how AI systems represent and process market information.
\end{abstract}

\section{Introduction}

The rapid expansion of algorithmic trading and AI-driven financial decision systems has raised growing regulatory concerns about ``model monoculture risk'' within financial infrastructures, particularly as firms converge on shared foundation models and common feature-extraction pipelines. Evidence indicates that highly synchronized algorithmic behavior can amplify market volatility and flash crashes.

This paper asks whether similarity in how AI agents \emph{represent} market states---as distinct from similarity in their realized predictions---increases systemic fragility in financial markets. More specifically, can representation homogeneity compress the effective space of forecast disagreement under stress, even when agents appear diverse in normal times?

To answer these questions, this paper develops a structural multi-agent framework that combines a representation-based AI microfoundation with a tractable reduced-form trading rule. The objective is not to replicate every engineering detail of deployed AI systems, but to construct a class of agents that is sufficiently rich to capture meaningful differences in information encoding while remaining sufficiently comparable to quantify representation overlap across agents. This makes it possible to study how structural similarity in the feature space through which agents interpret market states propagates into synchronized beliefs, correlated positions, and aggregate market instability.

A central contribution of the paper is the formal separation of \emph{representation homogeneity} from \emph{forecast overlap}. Two agents may produce similar return forecasts for transient reasons---correlated noise, similar recent experience---without sharing a common feature space. Conversely, agents that encode market states through similar representations are structurally predisposed to emphasize the same signals, neglect the same signals, and commit similar forecast errors, especially under stress. We formalize this distinction using function-space distances and show that representation convergence can compress the bound on forecast disagreement even when readout parameters differ. The experimental design is structured to test this specific claim: by constructing agent populations that are matched on forecast overlap but differ in representation distance, we can isolate the independent contribution of representation homogeneity to systemic fragility.

Risk-aversion heterogeneity and learning-rate heterogeneity enter the framework as static, agent-specific parameters whose cross-sectional distributions are varied as control dimensions in a factorial design. This allows us to verify that observed representation effects are robust to variation in downstream decision-making characteristics, without treating these dimensions as co-equal objects of investigation.

\section{Conceptual Framework}

The objective of the framework is to study \emph{representation homogeneity} as a structural object, rather than to classify traders into coarse strategy types. In heterogeneous-agent finance, aggregate outcomes depend not only on average behavior but also on the extent to which individual decision rules overlap \cite{tesfatsion2002agent,brock1998heterogeneous,hommes2006heterogeneous,lebaron2006agent}. This perspective is especially relevant in AI-mediated trading environments, where agents may differ in parameters while remaining highly similar in how they encode market states.

We represent agent \(i\) as a two-stage decision system,
\begin{equation}
\mathrm{Agent}_i = (f_i,g_i),
\end{equation}
where \(f_i\) maps the observable market state into a return forecast and \(g_i\) maps that forecast into a portfolio decision. Let
\[
S_t \in \mathcal{S} \subseteq \mathbb{R}^{K_s}
\]
denote the publicly observable market state at time \(t\). The state vector may include prices, returns, order-flow statistics, volatility measures, and other variables available prior to trading.

Agent \(i\)'s forecast of the next-period return is written abstractly as
\begin{equation}
\hat r_{i,t} = f_i(S_t),
\label{eq:abstract_forecast}
\end{equation}
and the resulting portfolio decision is
\begin{equation}
D_{i,t} = g_i(\hat r_{i,t},\sigma_t,D_{i,t-1}),
\label{eq:abstract_trading_rule}
\end{equation}
where \(\sigma_t\) denotes the volatility signal entering the trading rule and \(D_{i,t-1}\) is the inherited position. This notation is deliberately general. It accommodates statistical forecasting rules, machine-learning predictors, and adaptive trading heuristics while preserving a common decision architecture \cite{arthur1994inductive,brock1998heterogeneous,lebaron2006agent}.

Our interest, however, is not in this abstract decomposition itself, but in the structural layers through which homogeneity can arise. In the model developed below, agent \(i\)'s forecasting rule is represented as
\[
f_i(S)=\theta_i^\top h_i(S),
\qquad
h_i(S)=\phi(W_i S),
\]
where \(h_i:\mathcal{S}\to\mathbb{R}^K\) is the representation map, \(W_i\) is the representation matrix, \(\phi(\cdot)\) is an elementwise activation function, and \(\theta_i\) is the readout vector.

This decomposition separates three distinct objects. The \emph{representation layer} governs how raw market states are encoded into features. The \emph{forecast layer} governs how encoded features are mapped into return expectations. The \emph{risk-control layer} governs how those forecasts are translated into positions through risk tolerance, leverage limits, and adjustment frictions. Although all three layers can in principle contribute to agent similarity, the central premise of the paper is that \emph{upstream} similarity in representation is qualitatively different from, and potentially more destabilizing than, downstream similarity in forecasts or risk parameters. This is because representation homogeneity creates structural overlap in how agents \emph{perceive} market states, which can remain latent during normal times but become acutely consequential under stress.

This distinction is economically important because parameter similarity is not, by itself, a meaningful measure of overlap. Neural representations are not uniquely identified at the parameter level. Two agents may have different matrices \(W_i\) and \(W_j\) while generating similar feature maps over economically relevant states, and conversely, numerically similar parameter matrices may induce materially different encodings under nonlinear activation. For this reason, homogeneity is measured primarily in function space rather than raw parameter space.

\subsection{Structural Measures of Homogeneity}

We begin with the representation layer. For agents \(i\) and \(j\), define the pairwise representation distance as the \(L^2(\mu)\) distance between their feature maps:
\begin{equation}
d_{\mathrm{repr}}(i,j)
=
\left(
\int_{\mathcal{S}}
\|h_i(S)-h_j(S)\|_2^2
\, d\mu(S)
\right)^{1/2}.
\label{eq:repr_distance}
\end{equation}
Here \(\mu\) is a reference distribution over economically relevant market states. This distance measures whether two agents encode the same market state into similar feature vectors.

At the forecast layer, define
\begin{equation}
d_{\mathrm{forecast}}(i,j)
=
\left(
\int_{\mathcal{S}}
\bigl(f_i(S)-f_j(S)\bigr)^2
\, d\mu(S)
\right)^{1/2}.
\label{eq:forecast_distance}
\end{equation}
This object compares prediction rules directly. Two agents may generate similar forecasts even when their internal representations differ, and similar representations do not imply identical forecasts when readout vectors differ. Distinguishing \(d_{\mathrm{repr}}\) from \(d_{\mathrm{forecast}}\) is therefore essential for separating structural overlap in information processing from overlap in realized predictions.

At the risk-control layer, let \(\varpi_i\in\mathbb{R}^{K_R}\) denote the vector of parameters governing agent \(i\)'s position-sizing rule, including, for example, risk aversion, leverage limits, and adjustment frictions. We define the pairwise risk-control distance as
\begin{equation}
d_{\mathrm{risk}}(i,j)
=
\|\varpi_i-\varpi_j\|_{W},
\label{eq:risk_distance}
\end{equation}
where \(\|\cdot\|_{W}\) is a weighted norm chosen so that the resulting distances are comparable across components.

These three objects are the primitive measures used throughout the paper. When a scalar summary is useful, we define the composite distance
\begin{equation}
\mathcal{D}(i,j)
=
w_1 d_{\mathrm{repr}}(i,j)
+
w_2 d_{\mathrm{forecast}}(i,j)
+
w_3 d_{\mathrm{risk}}(i,j),
\label{eq:heterogeneity_new}
\end{equation}
where \(w_1,w_2,w_3\ge 0\) and \(w_1+w_2+w_3=1\). The associated homogeneity index is
\begin{equation}
\mathcal{H}(i,j)=1-\mathcal{D}(i,j),
\label{eq:homogeneity_new}
\end{equation}
and average market homogeneity is
\begin{equation}
\overline{\mathcal{H}}
=
\frac{2}{N(N-1)}
\sum_{1\le i<j\le N}
\mathcal{H}(i,j).
\label{eq:avg_homogeneity_new}
\end{equation}
In the analysis below, the weights \(w_1,w_2,w_3\) are treated as normative baseline settings for comparative statics and robustness checks rather than as structurally estimated parameters. The composite index \(\mathcal{H}(i,j)\) is used as a summary statistic, not as a substitute for the layer-specific distances.

\subsection{Operational Choice of the Reference Distribution}

The distances in \eqref{eq:repr_distance} and \eqref{eq:forecast_distance} depend on the reference distribution \(\mu\). Operationally, three choices are useful.

A first approach uses the empirical distribution of realized states \(\{S_t\}_{t=1}^T\) observed during calibration or simulation:
\begin{equation}
d_{\mathrm{repr}}(i,j)
=
\left(
\frac{1}{T}
\sum_{t=1}^{T}
\|\phi(W_i S_t)-\phi(W_j S_t)\|_2^2
\right)^{1/2}.
\label{eq:repr_empirical}
\end{equation}
This evaluates similarity over the region of the state space actually visited by the market, but it may underweight rare stress states.

A second approach evaluates distances over a predetermined benchmark grid
\[
\mathcal{S}_M=\{S^{(1)},\dots,S^{(M)}\},
\]
yielding
\begin{equation}
d_{\mathrm{repr}}(i,j)
=
\left(
\frac{1}{M}
\sum_{m=1}^{M}
\|\phi(W_i S^{(m)})-\phi(W_j S^{(m)})\|_2^2
\right)^{1/2}.
\label{eq:repr_grid}
\end{equation}
This allows explicit coverage of extreme states, but requires choosing a grid that is economically meaningful.

A third approach uses a weighted reference measure \(\mu=\omega(S)\mu_0(S)\), where \(\omega(S)\) places greater mass on states of particular interest, such as stress regimes:
\begin{equation}
d_{\mathrm{repr}}(i,j)
=
\left(
\int_{\mathcal{S}}
\|h_i(S)-h_j(S)\|_2^2 \,\omega(S)
\, d\mu_0(S)
\right)^{1/2}.
\label{eq:repr_weighted}
\end{equation}
This formulation is especially useful when the object of interest is not average similarity over normal times but overlap under tail conditions. In the analysis below, the choice of \(\mu\) is treated as part of the measurement design rather than as a primitive feature of the model.

\subsection{Why Representation and Forecast Homogeneity Are Distinct}

Representation similarity and forecast similarity are related, but they are not equivalent. Under mild regularity conditions, one can bound the second object using the first without collapsing the two concepts.

Assume that \(\phi(\cdot)\) is Lipschitz continuous, that \(\|\theta_i\|\le B_\theta\), and that the relevant state space is bounded. Then for any state \(S\),
\begin{align}
|f_i(S)-f_j(S)|
&=
|\theta_i^\top h_i(S)-\theta_j^\top h_j(S)| \\
&\le
\|\theta_i-\theta_j\|\,\|h_i(S)\|
+
\|\theta_j\|\,\|h_i(S)-h_j(S)\|.
\end{align}
Integrating with respect to \(\mu\) yields
\begin{equation}
d_{\mathrm{forecast}}(i,j)
\le
C_1 \|\theta_i-\theta_j\|
+
C_2 d_{\mathrm{repr}}(i,j),
\label{eq:forecast_bound}
\end{equation}
for constants \(C_1,C_2>0\) determined by norm bounds and the choice of \(\mu\). Equation \eqref{eq:forecast_bound} shows that representation convergence can compress forecast heterogeneity, but it does not eliminate the role of readout heterogeneity.

Importantly, \(d_{\mathrm{forecast}}(i,j)\) is an integrated distance defined over the full reference measure \(\mu\), whereas the theoretical results developed below focus on the state-conditional cross-sectional disagreement \(\mathcal{V}_f(S)\). This distinction is important: the former summarizes average overlap across states, while the latter isolates how representation convergence may contract the effective space of disagreement in economically important states, especially under stress.

\section{Model}

To study how representation homogeneity propagates into market-wide instability, we construct a discrete-time multi-agent market with four linked components: (i) a reduced-form pricing block with inventory-sensitive liquidity provision, (ii) a representation-based forecasting architecture, (iii) a risk-controlled portfolio choice rule, and (iv) an adaptive learning rule with an explicit prediction target and internally consistent timing.

Time is indexed by \(t=0,1,2,\dots\). Each period follows the sequence
\[
\text{state observation}
\;\rightarrow\;
\text{forecast formation}
\;\rightarrow\;
\text{portfolio choice}
\;\rightarrow\;
\text{price formation}
\;\rightarrow\;
\text{learning update}.
\]
This timing implies that agents form forecasts using information available at time \(t\), choose positions within period \(t\), and update their forecasting rules only after the next-period realization becomes observable.

\subsection{State Variables and Market Timeline}

Let
\[
S_t \in \mathcal{S}\subseteq \mathbb{R}^{K_s}
\]
denote the observable market state at the beginning of period \(t\). The state vector may include lagged returns, prices, realized volatility measures, and order-flow statistics. Crucially, \(S_t\) is measurable with respect to the information set available before agents choose their period-\(t\) trades.

A single period unfolds in five steps. First, agents observe the public state \(S_t\). Second, each agent maps \(S_t\) into features and forms a forecast of the next-period return. Third, agents choose their period-\(t\) positions \(D_{i,t}\), thereby generating order flow. Fourth, a reduced-form market maker absorbs aggregate order flow and sets the end-of-period price \(P_t\). Fifth, after the subsequent return realization is observed, agents update their forecasting readouts using the realized prediction error. A forecast formed at time \(t\) is therefore evaluated against the realized return from \(t\) to \(t+1\), not against a contemporaneous return already embedded in \(S_t\).

\subsection{Price Dynamics and Inventory-Sensitive Liquidity Provision}

We model a single risky asset traded against a reduced-form, risk-averse market maker. Let \(D_{i,t}\) denote agent \(i\)'s position at the end of period \(t\), and define the agent's order flow as
\begin{equation}
\Delta D_{i,t} = D_{i,t} - D_{i,t-1}.
\label{eq:flow_agent_rewrite}
\end{equation}
Aggregate net order flow in period \(t\) is
\begin{equation}
\Delta Q_t = \sum_{i=1}^{N} \Delta D_{i,t}.
\label{eq:flow_aggregate_rewrite}
\end{equation}

To eliminate mechanical predictability from fundamentals, we assume that the asset's fundamental value follows a random walk:
\begin{equation}
F_t = F_{t-1} + \epsilon_t,
\qquad
\epsilon_t \sim \mathcal{N}(0,\sigma_\epsilon^2).
\label{eq:fundamental_rewrite}
\end{equation}

Let \(I_t\) denote the market maker's inventory after absorbing period-\(t\) order flow:
\begin{equation}
I_t = I_{t-1} - \Delta Q_t.
\label{eq:inventory_update_rewrite}
\end{equation}
The end-of-period transaction price is
\begin{equation}
P_t = F_t + \lambda_t \Delta Q_t - \psi_t I_t,
\label{eq:price_level_rewrite}
\end{equation}
where \(\lambda_t\) is a temporary order-flow impact coefficient and \(\psi_t\) is an inventory-holding premium.

To capture endogenous liquidity withdrawal under stress while preserving boundedness, we let both coefficients depend nonlinearly on lagged dealer inventory:
\begin{align}
\lambda_t
&=
\lambda_0
\left(
1+
\frac{\alpha_\lambda I_{t-1}^2}{1+\beta_\lambda I_{t-1}^2}
\right),
\label{eq:lambda_rewrite}
\\
\psi_t
&=
\psi_0
\left(
1+
\frac{\alpha_\psi I_{t-1}^2}{1+\beta_\psi I_{t-1}^2}
\right).
\label{eq:psi_rewrite}
\end{align}
Here, \(\lambda_0,\psi_0>0\) are baseline liquidity-friction parameters, \(\alpha_\lambda,\alpha_\psi>0\) govern stress amplification, and \(\beta_\lambda,\beta_\psi>0\) ensure bounded responses.

The realized return from \(t\) to \(t+1\) is
\begin{equation}
r_{t+1} = P_{t+1}-P_t.
\label{eq:return_def_rewrite}
\end{equation}
Substituting \eqref{eq:fundamental_rewrite}--\eqref{eq:price_level_rewrite} and using \eqref{eq:inventory_update_rewrite} yields
\begin{equation}
r_{t+1}
=
\epsilon_{t+1}
+
(\lambda_{t+1}+\psi_{t+1})\Delta Q_{t+1}
-
\lambda_t \Delta Q_t
-
(\psi_{t+1}-\psi_t)I_t.
\label{eq:return_dynamics_rewrite}
\end{equation}
Equation \eqref{eq:return_dynamics_rewrite} shows that next-period returns combine exogenous news, current order-flow pressure, lagged impact reversal, and inventory revaluation induced by time-varying dealer stress.

\subsection{Representation-Based Forecasting Architecture}

At the beginning of period \(t\), agent \(i\) observes \(S_t\) and transforms it into a \(K\)-dimensional feature vector:
\begin{equation}
X_{i,t} = \phi(W_{i,t} S_t),
\label{eq:representation_rewrite}
\end{equation}
where \(W_{i,t}\in\mathbb{R}^{K\times K_s}\) is the agent's representation matrix and \(\phi(\cdot)\) is an elementwise nonlinear activation function.

Conditional on this representation, the agent forms a forecast of the next-period return:
\begin{equation}
\hat r_{i,t} = \theta_{i,t}^{\top} X_{i,t},
\label{eq:forecast_rewrite}
\end{equation}
where \(\theta_{i,t}\in\mathbb{R}^{K}\) is the readout vector. By construction, \(\hat r_{i,t}\) is a forecast for \(r_{t+1}\), not for a contemporaneous realized return.

\subsection{Perceived Trading Frictions and Local Slippage Estimation}

Agents do not observe the market maker's true impact coefficient \(\lambda_t\), aggregate order flow \(\Delta Q_t\), or inventory \(I_t\). Instead, each agent maintains a local estimate of execution frictions based on its own realized trading experience.

We model agent \(i\)'s perceived impact coefficient as an exponentially weighted moving average of realized slippage, conceptually related to an Amihud-style local illiquidity proxy:
\begin{equation}
\hat\lambda_{i,t}
=
(1-\rho)\hat\lambda_{i,t-1}
+
\rho
\left|
\frac{r_t}{\max\{|\Delta D_{i,t-1}|,\varepsilon\}}
\right|,
\qquad
\rho\in(0,1),\ \varepsilon>0.
\label{eq:lambda_hat_rewrite}
\end{equation}
The regularization constant \(\varepsilon\) prevents the estimator from exploding when an agent submits negligible volume. This object is interpreted as a reduced-form local slippage indicator rather than as a consistent estimate of the true aggregate impact coefficient.

Because the forecast \(\hat r_{i,t}\) is used to guide a position change \(\Delta D_{i,t}\), the agent evaluates forecast performance using an execution-adjusted next-period return:
\begin{equation}
\tilde r_{i,t+1}
=
r_{t+1}
-
\hat\lambda_{i,t}\, |\Delta D_{i,t}|.
\label{eq:effective_return_rewrite}
\end{equation}
This target penalizes aggressive turnover under perceived illiquidity and prevents the agent from mechanically attributing all realized returns to predictive success.

\subsection{Risk-Controlled Portfolio Choice}

After forming the forecast \(\hat r_{i,t}\), agent \(i\) chooses its period-\(t\) position \(D_{i,t}\). To capture the interaction between expected returns, risk control, and adjustment frictions, we assume that the agent solves
\begin{equation}
U_{i,t}
=
D_{i,t}\hat r_{i,t}
-
\frac{\gamma_i}{2}\hat\sigma_t^2 D_{i,t}^2
-
\frac{\kappa+\hat\lambda_{i,t}}{2}(D_{i,t}-D_{i,t-1})^2,
\label{eq:utility_rewrite}
\end{equation}
where \(\gamma_i>0\) is a static, agent-specific risk aversion parameter, \(\kappa>0\) is a baseline position-adjustment friction, and \(\hat\lambda_{i,t}\) is the local slippage estimate defined in \eqref{eq:lambda_hat_rewrite}.

The volatility signal entering the objective is endogenous and updated from realized returns using a centered exponential moving average:
\begin{align}
\hat\mu_t
&=
(1-\beta)\hat\mu_{t-1}
+
\beta r_t,
\label{eq:mu_rewrite}
\\
\hat\sigma_t^2
&=
(1-\beta)\hat\sigma_{t-1}^2
+
\beta (r_t-\hat\mu_t)^2,
\label{eq:sigma_rewrite}
\end{align}
with \(\beta\in(0,1)\). This specification allows prolonged smooth price trends to compress perceived risk even while positions are gradually accumulating.

The first-order condition for \eqref{eq:utility_rewrite} yields the unconstrained target position:
\begin{equation}
D_{i,t}^{*}
=
\frac{
\hat r_{i,t}
+
(\kappa+\hat\lambda_{i,t})D_{i,t-1}
}{
\gamma_i\hat\sigma_t^2
+
\kappa+\hat\lambda_{i,t}
}.
\label{eq:unconstrained_rewrite}
\end{equation}
Equation \eqref{eq:unconstrained_rewrite} is a partial-adjustment portfolio rule. The numerator captures both directional conviction and position persistence, while the denominator scales responsiveness by perceived risk and trading frictions.

Imposing a hard capital or leverage limit \(\bar D_i>0\), the actual position is
\begin{equation}
D_{i,t}
=
\operatorname{sgn}(D_{i,t}^{*})
\min\left\{
|D_{i,t}^{*}|,\bar D_i
\right\}.
\label{eq:final_position_rewrite}
\end{equation}

\subsection{Learning Dynamics and Representation Drift}

We assume that the representation matrix \(W_{i,t}\) evolves slowly relative to the high-frequency trading horizon, so that within a short execution window the primary adaptive margin is the readout vector \(\theta_{i,t}\). Learning therefore takes the form of stochastic gradient descent on squared prediction error using the execution-adjusted target \(\tilde r_{i,t+1}\).

After \(r_{t+1}\) is realized, agent \(i\) updates its readout according to
\begin{equation}
\theta_{i,t+1}
=
\theta_{i,t}
+
\eta_{\theta,i}
\bigl(
\tilde r_{i,t+1}-\hat r_{i,t}
\bigr)X_{i,t},
\label{eq:theta_update_rewrite}
\end{equation}
where \(\eta_{\theta,i}>0\) is the agent's fixed learning rate. Substituting \eqref{eq:effective_return_rewrite}, the update can equivalently be written as
\begin{equation}
\theta_{i,t+1}
=
\theta_{i,t}
+
\eta_{\theta,i}
\Bigl(
r_{t+1}
-
\hat\lambda_{i,t}|\Delta D_{i,t}|
-
\hat r_{i,t}
\Bigr)X_{i,t}.
\label{eq:theta_update_expanded_rewrite}
\end{equation}
This timing is internally consistent: \(X_{i,t}\) and \(\hat r_{i,t}\) are computed from information available at time \(t\), \(\Delta D_{i,t}\) is the trade implied by that forecast, \(r_{t+1}\) is the realized next-period outcome, and the forecast error is evaluated only after the outcome becomes observable.

To capture gradual convergence in the underlying feature space, we allow individual representation matrices to evolve according to a matrix-valued Ornstein--Uhlenbeck process:
\begin{equation}
dW_{i,t}
=
\nu_W \bigl(W_{\mathrm{base},t}-W_{i,t}\bigr)dt
+
\sigma_W d\mathcal{B}_{i,t}^{W},
\label{eq:ou_rewrite}
\end{equation}
where \(\nu_W>0\) governs the speed of convergence toward a common benchmark representation, \(\sigma_W>0\) governs idiosyncratic exploratory drift, and \(W_{\mathrm{base},t}\) denotes a shared foundation representation following a common external updating process. This component allows the model to capture the slow-moving process by which firms adopt common AI infrastructure, while the readout layer adapts at high frequency to realized market conditions.

Note that the risk-aversion parameter \(\gamma_i\) and the learning rate \(\eta_{\theta,i}\) are both static and agent-specific: they are drawn from cross-sectional distributions at initialization and remain fixed throughout the simulation. The only dynamic convergence mechanism in the model operates at the representation level through \eqref{eq:ou_rewrite}. This design choice reflects the empirical observation that the most consequential source of structural convergence in AI-driven finance is the adoption of common foundation models and shared feature-extraction pipelines, rather than the standardization of risk preferences or learning algorithms per se.

\section{Structural Calibration via Simulated Method of Moments}

To avoid arbitrary parameterization of the market microstructure, we calibrate the baseline pricing and liquidity parameters using the Simulated Method of Moments (SMM) \cite{duffie1993simulated}. The calibration targets the reduced-form pricing block---specifically, the liquidity provision mechanism and fundamental volatility---rather than the agent-level decision parameters.

The structural parameter vector entering the SMM is
\[
\Theta = \{\lambda_0, \alpha_\lambda, \beta_\lambda, \psi_0, \alpha_\psi, \beta_\psi, \kappa, \sigma_\epsilon\}.
\]
This vector governs the market maker's pricing rule, the nonlinear liquidity response, and the scale of exogenous fundamental innovation.

Agent-level parameters---specifically, the risk-aversion distribution \(\gamma_i \sim \text{LogNormal}(\mu_\gamma, \sigma_\gamma)\) and the learning-rate distribution \(\eta_{\theta,i} \sim \text{LogNormal}(\mu_\eta, \sigma_\eta)\)---are treated as exogenous background heterogeneity. Their distributional parameters \((\mu_\gamma, \sigma_\gamma)\) and \((\mu_\eta, \sigma_\eta)\) are set to empirically reasonable values drawn from existing literature on institutional trading behavior and held fixed during calibration. These distributions are subsequently varied in the experimental design (Section~5) as control dimensions, not as calibration targets. In the baseline calibration, both distributions are set to moderate heterogeneity levels; sensitivity to alternative baseline means is assessed in the robustness analysis.

To improve identification of the nonlinear fragility parameters (\(\alpha\)) and the capacity saturation parameters (\(\beta\)), we expand the SMM target vector to include conditional tail moments at multiple volatility thresholds.

\begin{table}[ht]
\centering
\caption{SMM Target Moments and Identification Strategy}
\label{tab:smm_moments}
\begin{tabular}{@{}ll@{}}
\toprule
\textbf{Target Empirical Moment} & \textbf{Primary Structural Role} \\ \midrule
Annualized Volatility & Exogenous fundamental risk ($\sigma_\epsilon$) \\
Unconditional Price Impact & Baseline temporary impact ($\lambda_0$) \\
\textbf{Conditional Tail Price Impact (Top 5\% Vol)} & \textbf{Liquidity fragility parameter ($\alpha_\lambda$)} \\
\textbf{Extreme Tail Price Impact (Top 1\% Vol)} & \textbf{Liquidity capacity saturation ($\beta_\lambda$)} \\
Return Autocorrelation (Lag 1) & Baseline inventory premium ($\psi_0$) \\
\textbf{Conditional Tail Return Autocorrelation (Top 5\%)} & \textbf{Inventory fragility parameter ($\alpha_\psi$)} \\
\textbf{Extreme Tail Return Autocorrelation (Top 1\%)} & \textbf{Inventory capacity saturation ($\beta_\psi$)} \\
Order Flow Autocorrelation (Lag 1) & Adjustment friction / Stickiness ($\kappa$) \\
\bottomrule
\end{tabular}
\end{table}

The calibration proceeds in two stages. In the first stage, a global search over the parameter space identifies candidate regions using Sobol quasi-random sequences. In the second stage, local optimization from the best candidates refines the estimates. Bootstrap resampling of the empirical moments provides confidence intervals. The calibrated baseline market serves as the common environment for all subsequent experiments: comparative statics on representation homogeneity are conducted on top of this calibrated pricing block.

\section{Experimental Design: Identification of Representation Homogeneity Effects}

The experimental design is structured to answer a specific identification question: \emph{does representation homogeneity increase systemic fragility, and does this effect persist after controlling for variation in risk-aversion and learning-rate distributions?}

Unlike many agent-based studies that vary multiple structural parameters simultaneously, we adopt a controlled factorial design in which representation homogeneity serves as the primary treatment variable, while risk-control and learning-rate heterogeneity serve as control dimensions whose cross-sectional distributions are independently varied to verify robustness.

\subsection{Treatment and Control Dimensions}

\paragraph{Primary Treatment: Representation Homogeneity (\(H_W\))}

Representation homogeneity measures similarity in the feature-extraction layer \(h_i(S)=\phi(W_i S)\). This is the central object of investigation.

\begin{itemize}
\item \textbf{Low Homogeneity} (\(H_W=0\)): Representation matrices are drawn from a wide distribution \(W_i \sim \mathcal{D}_W^{\text{wide}}\), producing large average pairwise representation distances.
\item \textbf{High Homogeneity} (\(H_W=1\)): Representation matrices are tightly clustered around a common benchmark \(W_i \sim \mathcal{D}_W^{\text{tight}}\), producing small average pairwise representation distances.
\end{itemize}

The mean scale of \(W_i\) is preserved across regimes so that the treatment modifies only the average pairwise representation distance
\[
\overline d_{\mathrm{repr}}
=
\frac{2}{N(N-1)}
\sum_{i<j} d_{\mathrm{repr}}(i,j).
\]

\paragraph{Control Dimension: Risk-Aversion Distribution (\(H_\gamma\))}

Risk-aversion parameters are drawn from
\[
\gamma_i \sim \text{LogNormal}(\mu_\gamma,\sigma_\gamma)
\]
and remain fixed throughout the simulation. The two regimes differ only in cross-sectional dispersion:

\begin{itemize}
\item \(H_\gamma=0\): \(\sigma_\gamma = \sigma_\gamma^{\text{high}}\) (heterogeneous risk preferences)
\item \(H_\gamma=1\): \(\sigma_\gamma = \sigma_\gamma^{\text{low}}\) (uniform risk preferences)
\end{itemize}

while maintaining \(\mathbb{E}[\gamma_i] = \bar\gamma\) in both regimes. This dimension serves as a control: it verifies that representation effects are not confounded by variation in downstream risk-taking behavior.

\paragraph{Control Dimension: Learning-Rate Distribution (\(H_\eta\))}

Learning rates are drawn from
\[
\eta_{\theta,i} \sim \text{LogNormal}(\mu_\eta,\sigma_\eta)
\]
and remain fixed throughout the simulation. The two regimes are:

\begin{itemize}
\item \(H_\eta=0\): \(\sigma_\eta = \sigma_\eta^{\text{high}}\) (heterogeneous learning speeds)
\item \(H_\eta=1\): \(\sigma_\eta = \sigma_\eta^{\text{low}}\) (uniform learning speeds)
\end{itemize}

while holding \(\mathbb{E}[\eta_{\theta,i}] = \bar\eta\) fixed.

\subsection{Factorial Structure}

The three dimensions produce a \(2\times 2\times 2\) factorial design:
\[
(H_W,H_\gamma,H_\eta)\in\{0,1\}^3.
\]
This yields eight experimental configurations. The factorial structure is used not to identify ``which dimension matters most,'' but to estimate the representation homogeneity effect \(\beta_W\) while controlling for potential confounds from risk-aversion and learning-rate distributions.

A key identification requirement is that treatments alter the \emph{dispersion} of agent characteristics rather than their mean levels. Otherwise, treatment effects could reflect shifts in average behavior rather than structural homogeneity.

\subsection{Monte Carlo Simulation Procedure}

For each experimental configuration \(c\in\{1,\dots,8\}\), we perform \(M\) independent Monte Carlo simulations of the market model. Let \(Y_{c,m}\) denote a systemic outcome statistic from simulation \(m\) under configuration \(c\). The estimated outcome is
\[
\bar Y_c = \frac{1}{M} \sum_{m=1}^M Y_{c,m}.
\]
Standard errors are computed across Monte Carlo replications.

\subsection{Mechanism Decomposition}

To understand how representation homogeneity propagates through the system, we evaluate outcomes at multiple layers of the market dynamics.

\paragraph{Representation Synchronization}

Average representation distance:
\[
\overline d_{\mathrm{repr}} = \frac{2}{N(N-1)} \sum_{i<j} d_{\mathrm{repr}}(i,j).
\]

\paragraph{Forecast Synchronization}

Cross-agent forecast correlation:
\[
\rho^{\mathrm{forecast}} = \text{Corr}(\hat r_{i,t},\hat r_{j,t}).
\]

\paragraph{Position Synchronization}

Cross-agent position correlation:
\[
\rho^{\mathrm{position}} = \text{Corr}(D_{i,t},D_{j,t}).
\]

\paragraph{Order Flow Concentration}

Aggregate order-flow concentration is measured by
\[
C_t = \frac{|\sum_i \Delta D_{i,t}|}{\sum_i |\Delta D_{i,t}|}.
\]
Higher values indicate more directional market-wide trading.

\paragraph{Liquidity Stress}

Liquidity pressure is measured using the average price impact coefficient \(\lambda_t\), the average inventory premium \(\psi_t\), and dealer inventory stress \(|I_t|\).

\paragraph{Tail Risk}

Systemic outcomes include crash frequency (proportion of returns exceeding a threshold multiple of unconditional volatility), extreme return quantiles (1\% and 5\% Value-at-Risk), and maximum drawdown.

These measures allow us to trace the mechanism
\[
\text{representation similarity}
\rightarrow
\text{forecast synchronization}
\rightarrow
\text{position synchronization}
\rightarrow
\text{order-flow concentration}
\rightarrow
\text{liquidity stress}
\rightarrow
\text{tail risk}.
\]

\subsection{Factorial Effect Estimation}

To quantify treatment effects, we estimate a factorial regression across experimental configurations:
\begin{align}
Y
&=
\alpha
+
\beta_W H_W
+
\beta_\gamma H_\gamma
+
\beta_\eta H_\eta
+
\beta_{W\gamma} H_W H_\gamma
+
\beta_{W\eta} H_W H_\eta
+
\beta_{\gamma\eta} H_\gamma H_\eta
+
\beta_{W\gamma\eta} H_W H_\gamma H_\eta
+
\varepsilon.
\end{align}

The primary object of interest is \(\beta_W\), which measures the direct effect of representation homogeneity on the outcome variable. The coefficients \(\beta_\gamma\) and \(\beta_\eta\) serve as controls, verifying that observed effects are not driven by variation in risk-aversion or learning-rate distributions. Interaction terms \(\beta_{W\gamma}\) and \(\beta_{W\eta}\) are reported in the appendix as secondary findings, capturing whether the representation effect is amplified when downstream decision parameters are also more uniform.

\subsection{Single-Factor Representation Scan}

In addition to the factorial design, we conduct a continuous single-factor experiment that varies only representation dispersion while holding risk-aversion and learning-rate distributions fixed at their heterogeneous baseline levels (\(H_\gamma=0\), \(H_\eta=0\)). Representation dispersion is varied across 15 grid points from \(\sigma_W^{\text{wide}}\) to \(\sigma_W^{\text{tight}}\). This experiment provides transparent visual evidence of how representation distance relates to tail risk, and is used to identify potential threshold effects.

As a robustness check, the same scan is repeated under uniform risk-aversion and learning-rate distributions (\(H_\gamma=1\), \(H_\eta=1\)). If the two curves are close, this supports the conclusion that the representation effect is the dominant channel; if they diverge, the interaction merits further investigation.

\subsection{Operational Identification of the Critical Representation Threshold}

The hypothesis developed in Section~7 posits the existence of a critical representation-distance threshold \(\overline d_{\mathrm{repr}}^{\,\mathrm{crit}}\) below which tail risk rises sharply. The single-factor scan provides the natural setting for estimating an operational counterpart of this threshold.

Let
\[
\mathcal{F}(d)
=
\mathbb{E}\bigl[Y \mid \overline d_{\mathrm{repr}}=d\bigr],
\]
where \(Y\) denotes a tail-risk outcome such as crash frequency, extreme return quantiles, or peak execution cost. We define the estimated critical threshold
\[
\widehat{\overline d}_{\mathrm{repr}}^{\,\mathrm{crit}}
\]
as the change-point in the relationship between \(\overline d_{\mathrm{repr}}\) and \(\mathcal{F}(d)\), obtained from a segmented linear fit. Intuitively, this is the point at which marginal reductions in representation distance begin to produce disproportionately large increases in fragility. Spline-based maximum-curvature estimates and event-probability crossing rules are reported as robustness checks.

Because this threshold is outcome-specific by construction, we report estimates for multiple tail-risk measures and assess whether they cluster within a narrow range. Convergence of the estimated thresholds across different outcome variables would provide evidence that the critical transition reflects a structural regime shift in synchronization dynamics rather than an artifact of a particular fragility measure.

\subsection{Matched Forecast-Distance Identification Experiment}

A distinctive feature of this paper's framework is the formal separation of representation homogeneity from forecast overlap. To test this distinction empirically, we construct a matched-design experiment in which two groups of agents are selected to have similar integrated forecast distances \(d_{\mathrm{forecast}}\) but different representation distances \(d_{\mathrm{repr}}\).

\paragraph{Construction}

We generate a large number of candidate agent populations and compute both \(\bar d_{\mathrm{repr}}\) and \(\bar d_{\mathrm{forecast}}\) for each, using the empirical approximation of \(d_{\mathrm{forecast}}\) under the same reference distribution \(\mu\) defined in Section~2. Two groups are then selected:

\begin{itemize}
\item \textbf{Group A} (low \(d_{\mathrm{repr}}\), moderate \(d_{\mathrm{forecast}}\)): Agents encode market states through similar representations but maintain moderate forecast diversity through differences in readout vectors \(\theta_i\).
\item \textbf{Group B} (high \(d_{\mathrm{repr}}\), moderate \(d_{\mathrm{forecast}}\)): Agents encode market states through diverse representations, with forecast diversity at a similar level to Group A.
\end{itemize}

The matching criterion requires that the two groups' average forecast distances fall within a narrow band (difference \(<5\%\)), while their representation distances differ substantially.

If strict matching yields insufficient candidate populations (success rate \(<3\%\)), the design automatically transitions to a conditional binning approach: all candidate populations are sorted by \(\bar d_{\mathrm{forecast}}\) into bins, and within each bin, high- versus low-\(\bar d_{\mathrm{repr}}\) subgroups are compared.

\paragraph{Experimental Protocol}

Each group is simulated through two phases. In the \emph{calm phase}, the market operates under normal fundamental volatility for an extended period. In the \emph{stress phase}, a sequence of large fundamental shocks is injected. The hypothesis is that Groups A and B will behave similarly during the calm phase---since their forecast distances are matched---but that Group A will exhibit significantly greater synchronization, liquidity stress, and tail risk during the stress phase, because its agents share a common representational lens that compresses forecast disagreement under extreme states.

This experiment is designed to assess whether representation homogeneity carries incremental explanatory power for tail risk beyond what is captured by realized forecast overlap. A necessary condition for the matched comparison to be informative is that the two groups are well-matched during the calm phase; we therefore report calm-phase differences in forecast correlation, position correlation, and volatility as a diagnostic check before interpreting stress-phase results.

\subsection{Negative Control Experiments}

To ensure that observed fragility arises from representation-driven synchronization rather than from model-specific nonlinearities, we implement several negative controls:

\begin{enumerate}
\item \textbf{Constant liquidity}: fixing \(\lambda_t=\lambda_0\) and \(\psi_t=\psi_0\), which disables the nonlinear amplification channel.
\item \textbf{Shared signals without shared representations}: all agents observe the same \(S_t\) but use highly dispersed representation matrices.
\item \textbf{Linear representation benchmark}: \(h_i(S)=W_i S\), which removes the nonlinear activation.
\end{enumerate}

These controls verify that the tail risk amplification identified in the main experiments is driven by structural synchronization in the representation layer, not by mechanical features of the pricing equation.

\section{Representation Convergence Experiments}

The factorial design in Section~5 identifies the effect of representation homogeneity by independently manipulating the dispersion of agent characteristics at initialization. While this approach isolates structural effects in a controlled setting, real-world AI ecosystems are unlikely to transition abruptly from heterogeneous to homogeneous states. Instead, convergence typically occurs gradually as firms adopt common infrastructure, share training data pipelines, or deploy increasingly standardized foundation models.

To capture these transition dynamics, we now study environments in which representation matrices evolve stochastically over time toward a shared industry architecture. The objective of this section is not identification---which is already established in the factorial design---but rather to examine the \emph{temporal propagation} of synchronization and tail risk as representation homogeneity increases endogenously. The baseline convergence experiments treat the common benchmark \(W_{\mathrm{base}}\) as static; Section~8 relaxes this assumption by allowing the foundation model itself to evolve over time.

\subsection{Dynamic Evolution of Representation Matrices}

Representation matrices evolve according to the matrix-valued Ornstein--Uhlenbeck process introduced in Section~3:
\begin{equation}
dW_{i,t}
=
\nu_W (W_{\mathrm{base}} - W_{i,t})dt
+
\sigma_W d\mathcal{B}_{i,t}^{W},
\label{eq:ou_representation}
\end{equation}
where \(W_{\mathrm{base}}\) represents an industry-standard foundation model, \(\nu_W\) controls the speed of representation convergence, and \(\sigma_W\) governs exploratory deviations across agents.

Crucially, the risk-aversion parameters \(\gamma_i\) and learning rates \(\eta_{\theta,i}\) remain fixed at their initialized values throughout the convergence process. This isolates the effect of representation convergence from any simultaneous changes in risk preferences or learning behavior, consistent with the model specification in Section~3.

As \(t\to\infty\) or \(\nu_W\to\infty\), representation maps \(h_i(S)=\phi(W_{i,t}S)\) converge toward a common feature space, reducing the pairwise representation distance \(d_{\mathrm{repr}}(i,j)\) defined in Section~2.

\subsection{Dynamic Homogeneity Metrics}

To track convergence over time, we compute time-varying measures:

\begin{equation}
\overline d_{\mathrm{repr},t}
=
\frac{2}{N(N-1)}
\sum_{i<j}
d_{\mathrm{repr},t}(i,j),
\label{eq:dynamic_repr_distance}
\end{equation}
\begin{equation}
\rho^{\mathrm{forecast}}_t
=
\text{Corr}\big(\hat r_{i,t},\hat r_{j,t}\big),
\label{eq:dynamic_forecast_corr}
\end{equation}
\begin{equation}
\rho^{\mathrm{position}}_t
=
\text{Corr}\big(D_{i,t},D_{j,t}\big).
\label{eq:dynamic_position_corr}
\end{equation}

These statistics allow us to track the progressive tightening of the synchronization chain
\[
\text{representation convergence}
\rightarrow
\text{forecast synchronization}
\rightarrow
\text{position synchronization}.
\]

\subsection{Convergence Scenarios}

We simulate the dynamic system starting from a heterogeneous initial condition in which representation matrices are drawn from a wide cross-sectional distribution. Risk-aversion and learning-rate parameters are drawn from their baseline heterogeneous distributions and held fixed. The convergence process in equation~\eqref{eq:ou_representation} is then activated at different speeds:

\begin{table}[ht]
\centering
\begin{tabular}{@{}lll@{}}
\toprule
\textbf{Scenario} & \(\nu_W\) & \textbf{Interpretation} \\ \midrule
No convergence & 0 & Negative control \\
Slow convergence & 0.001 & Gradual industry standardization \\
Baseline convergence & 0.01 & Moderate adoption of common AI infrastructure \\
Fast convergence & 0.1 & Rapid industry consolidation \\
\bottomrule
\end{tabular}
\end{table}

For each scenario, we track representation distance \(\overline d_{\mathrm{repr},t}\), forecast correlation \(\rho^{\mathrm{forecast}}_t\), position correlation \(\rho^{\mathrm{position}}_t\), order-flow concentration, dealer inventory stress \(|I_t|\), and the liquidity coefficients \(\lambda_t\) and \(\psi_t\). A key object of interest is whether the time path of \(\overline d_{\mathrm{repr},t}\) crosses the critical threshold estimated in Section~5, and whether downstream fragility indicators deteriorate sharply at or near this crossing.

\section{Mechanism-Based Hypotheses}

\subsection{Hypothesis I: The Volatility Paradox and Hidden Leverage}

Driven by the partial-adjustment rule in Equation~\eqref{eq:unconstrained_rewrite}, we hypothesize that rising representation homogeneity will generate smoother collective trends in returns. According to Equation~\eqref{eq:sigma_rewrite}, these trends compress the centered perceived variance \(\hat\sigma_t^2\). Paradoxically, low perceived volatility then encourages agents to expand positions, while the stickiness term \(\kappa D_{i,t-1}\) allows leverage to accumulate gradually and opaquely.

This mechanism does not require that agents share identical risk preferences or learning rates. Even with heterogeneous \(\gamma_i\) and \(\eta_{\theta,i}\), agents whose representations converge will tend to forecast in the same direction, producing smoother aggregate returns and triggering the volatility-compression channel through the common perceived-volatility estimator.

\subsection{Hypothesis II: Critical Representation Threshold}

Representation homogeneity is not equivalent to temporary similarity in forecasts. Two agents may occasionally produce similar return forecasts even if they process information through very different feature spaces. By contrast, representation homogeneity means that agents interpret market states through similar informational lenses. As a result, they are likely to emphasize the same signals, neglect the same signals, and commit similar forecast errors.

The key theoretical implication concerns the \emph{state-conditional} nature of this overlap. The integrated forecast distance \(d_{\mathrm{forecast}}(i,j)\) averages disagreement over all states weighted by the reference measure \(\mu\). However, the systemic consequences of homogeneity depend primarily on disagreement in \emph{stress states}---states where coordinated position adjustments can overwhelm market maker capacity.

We hypothesize the existence of a positive critical level of average representation distance,
\[
\overline d_{\mathrm{repr}}^{\,\mathrm{crit}} > 0,
\]
such that when
\[
\overline d_{\mathrm{repr},t}
>
\overline d_{\mathrm{repr}}^{\,\mathrm{crit}},
\]
cross-agent disagreement remains sufficiently dispersed to stabilize aggregate order flow, whereas when
\[
\overline d_{\mathrm{repr},t}
\le
\overline d_{\mathrm{repr}}^{\,\mathrm{crit}},
\]
representation-driven synchronization becomes self-reinforcing, directional order flow intensifies, and tail risk rises sharply. The threshold is therefore interpreted as the point at which cross-sectional heterogeneity ceases to provide an effective buffer against common forecast errors.

That the critical threshold is strictly positive is important: fragility does not require agents to be identical. In the present framework, readout heterogeneity, idiosyncratic noise, leverage constraints, and adjustment frictions all continue to provide residual dispersion even when representations converge. The threshold emerges once this residual heterogeneity becomes too weak to buffer common directional errors induced by shared representations. Equivalently, the dangerous regime begins \emph{before} complete representational overlap---a distinction with direct implications for the timing of macroprudential intervention.

This hypothesis implies that matched populations with similar integrated forecast distances but lower representation distances should exhibit sharper stress-period synchronization and higher tail risk: by comparing two agent populations with similar integrated forecast distances but different representation distances, we can assess whether representation homogeneity predicts stress-period fragility beyond what forecast overlap alone would imply.

\subsection{Hypothesis III: Foundation-Model Drift and Common Blind Spots}

A distinguishing feature of representation homogeneity, relative to forecast correlation, is the creation of \emph{common blind spots}: systematic gaps in how agents perceive market states. Forecast correlation can be episodic and state-dependent, but representation homogeneity is structural. When agents share a common feature space, they are predisposed not only to agree when their shared features are informative, but also to \emph{miss the same signals} when their shared features are uninformative.

In the representation convergence experiments (Section~6), we examine whether this structural vulnerability persists even when the foundation model itself evolves over time. Under the non-stationary extension (Section~8), the feature manifold is constantly shifting. However, as long as the industry standardization speed \(\nu_W\) is sufficiently large, cross-agent representation distances still collapse toward zero. The common blind spots therefore move with the foundation model rather than being eliminated by its evolution.

% ============================================================
% SECTION 8 — Fixed: removed "Proposed" from title
% ============================================================

\section{Stress Tests and Robustness Checks}

To validate the fragility mechanisms identified above, we conduct a set of out-of-sample stress tests.

\subsection{Heavy-Tailed Exogenous Shocks}

Empirical high-frequency returns often exhibit jumps and heavy tails. We therefore replace the Gaussian innovation \(\epsilon_t\) with a jump-diffusion specification containing a symmetric \(\alpha\)-stable L\'evy component. The hypothesis is that markets with heterogeneous representations can absorb such shocks through staggered belief adjustments, whereas markets with homogeneous representations convert the same exogenous shock into an endogenous liquidity event.

\subsection{Asynchronous Learning and Stochastic Latency}

Institutional AI agents need not refresh their readout layers or execute trades at exactly the same instant. We therefore introduce stochastic asynchronous update schedules, modeled through independent Poisson clocks with agent-specific arrival rates. The hypothesis is that moderate latency dispersion delays the timing of synchronized deleveraging but does not eliminate the underlying instability once representation homogeneity is sufficiently high.

\subsection{Dynamic Foundation Model and Non-Stationary Environments}

In the representation convergence specification introduced in Section~6, the representation matrices of all agents converge toward a static industry-standard foundation model \(W_{\mathrm{base}}\). While this parsimonious specification isolates the effect of structural homogeneity, real-world foundation models are continuously retrained and fine-tuned on new data.

To ensure that our results are driven by the vanishing structural distance between agents rather than by a stagnant representation space, we introduce a non-stationary baseline model:
\begin{equation}
dW_{\mathrm{base},t} = \sigma_{\mathrm{base}}\, d\mathcal{Z}_t,
\end{equation}
where \(d\mathcal{Z}_t\) is a matrix of standard Brownian motions independent of the agents' idiosyncratic exploration terms, and \(\sigma_{\mathrm{base}}\) governs the innovation rate of the underlying AI architecture.

Consequently, the individual agent representation dynamics track this moving target:
\begin{equation}
dW_{i,t} = \nu_W \bigl(W_{\mathrm{base},t} - W_{i,t}\bigr)\, dt + \sigma_W\, d\mathcal{B}_{i,t}^W.
\end{equation}

Under this generalized specification, the feature manifold \(\phi(W_{\mathrm{base},t}S_t)\) is constantly shifting, allowing agents to continuously adapt to new market regimes. However, as the industry standardization speed \(\nu_W \rightarrow \infty\), cross-agent representation distances \(d_{\mathrm{repr}}(i,j)\) still collapse toward zero, so that agents process market states through increasingly similar---albeit evolving---informational lenses.

\section{Discussion and Mechanistic Interpretation}

\subsection{Resolving the Paradox: Local Adaptation vs.\ Global Collapse}

A natural question is how a tractable representation-readout architecture can explain far-from-equilibrium market crashes. The mechanism is bi-phasic. During the calm phase, adaptive readout updates in Equation~\eqref{eq:theta_update_expanded_rewrite} progressively align beliefs, while the price system and the volatility estimator compress perceived risk. During the crash phase, the nonlinear structure of the risk-control rule in Equation~\eqref{eq:unconstrained_rewrite} dominates. The sudden expansion of the volatility denominator clashes with sticky positions, pushing many agents across the same adjustment threshold at nearly the same time. Representation homogeneity matters because it makes this threshold crossing synchronized rather than staggered.

\subsection{Why Representation Homogeneity Is Especially Destabilizing}

A key implication of the representation-readout architecture is that representation homogeneity is not equivalent to temporary similarity in forecasts. Two agents may occasionally produce similar return forecasts even if they process information through very different feature spaces. By contrast, representation homogeneity means that agents interpret market states through similar informational lenses. As a result, they are likely to emphasize the same signals, neglect the same signals, and commit similar forecast errors. In systemic-risk terms, this creates common blind spots. Forecast correlation can be episodic, but representation homogeneity is structural: it increases the probability that agents will be wrong in the same direction at the same time.

The matched forecast-distance experiment in Section~5 is designed to test this claim directly. If two agent populations with similar forecast overlap but different representation distances exhibit markedly different behavior under stress, this would constitute evidence that representation homogeneity carries independent information about tail risk.

\subsection{Why the Critical Threshold Is Positive}

An important implication of Hypothesis~II is that the critical representation threshold \(\overline d_{\mathrm{repr}}^{\,\mathrm{crit}}\) is expected to be strictly positive. Fragility does not require agents to become identical. In the present framework, several residual sources of dispersion persist even under high representation homogeneity: readout heterogeneity from different \(\theta_i\), idiosyncratic exploration noise in the OU process, heterogeneous risk aversion \(\gamma_i\), diverse learning rates \(\eta_{\theta,i}\), and hard leverage constraints \(\bar D_i\). The critical threshold marks the point at which these residual buffers become insufficient to prevent common directional errors from cascading through the order-flow concentration channel.

This has a direct policy implication: the dangerous regime begins before complete representational overlap. A regulator monitoring representation distances need not wait until agents are nearly identical to intervene; the structurally relevant boundary lies at a positive distance that depends on the market's liquidity environment, friction parameters, and residual heterogeneity in downstream decision-making.

\subsection{Role of Risk-Control and Learning Heterogeneity}

The factorial design treats risk-aversion and learning-rate distributions as control dimensions rather than primary treatment variables. This does not imply that these characteristics are irrelevant to systemic stability. Rather, the experimental design is structured to test a more specific claim: that representation homogeneity is a \emph{necessary} upstream condition for the type of synchronized fragility studied here, and that its effects are robust to variation in downstream decision parameters.

If the factorial results show that \(\beta_W\) is large and significant while \(\beta_\gamma\) and \(\beta_\eta\) are small, this supports the conclusion that the primary source of synchronization risk lies in how agents encode information, not in how they translate forecasts into positions. If interaction effects \(\beta_{W\gamma}\) or \(\beta_{W\eta}\) are also significant, this would indicate that risk-control or learning uniformity can \emph{amplify} the representation effect, even if they do not independently generate it.

\subsection{Methodological Scope and Empirical Boundaries}

A central premise of this framework is the evaluation of financial markets under conditions of severe representation homogeneity among AI trading agents. We acknowledge that current financial ecosystems remain in a mixed state, characterized by a blend of diverse algorithmic systems, traditional institutional investors, and noise traders. Consequently, the high-representation-homogeneity configurations in the factorial design should be interpreted as boundary cases for macroprudential stress testing rather than as strictly contemporaneous empirical descriptions of current markets.

However, the methodological choice to simulate these boundary conditions is both deliberate and necessary. The objective of this structural approach is not to provide high-frequency point forecasts of current market states, but rather to conduct counterfactual analyses of fragility. By isolating the mechanics of representation-driven synchronization, the model functions as a controlled laboratory to identify the structural thresholds where a resilient heterogeneous ecology transitions into a fragile synchronized state.

Furthermore, while an ecosystem completely dominated by overlapping foundation models has not yet fully materialized, the underlying mechanisms of synchronized deleveraging have historical precedents. Localized episodes of structural convergence, such as the synchronized unwinding of quantitative models during historical liquidity events, provide empirical proxies for the dynamics captured here. By formalizing how hidden leverage accumulates during low-volatility regimes and violently unwinds during shocks, this framework equips regulators with the theoretical scaffolding required to monitor and mitigate representation monoculture risk before it manifests as extreme tail events in AI-dominated infrastructures.

\section{Policy Implications}

\begin{itemize}
\item \textbf{Monitoring representation-level concentration}: Regulators should develop tools to assess the degree of representational overlap among AI trading systems, going beyond the current focus on strategy classification or position reporting. Metrics analogous to \(d_{\mathrm{repr}}\) could provide early-warning indicators of structural synchronization risk. The finding that the critical representation threshold is strictly positive implies that intervention may be warranted well before agents become nearly identical.

\item \textbf{AI ecosystem design and upstream infrastructure}: Authorities should monitor concentration in upstream AI infrastructure, including common data vendors, shared feature-extraction pipelines, and heavily concentrated foundation-model providers. The results of this paper suggest that convergence at the representation level---where agents learn to \emph{see} markets similarly---may be more consequential for systemic stability than convergence in risk management parameters.

\item \textbf{Re-evaluating macroprudential paradigms}: While standardized risk-management requirements serve important prudential objectives, regulators should be aware that excessive standardization in how market participants \emph{process information} can create procyclical fragility. The key insight is that representation homogeneity can generate synchronized behavior even among agents with diverse risk preferences and trading strategies.
\end{itemize}

\section{Conclusion}

Similarity in how AI trading systems represent market states can propagate through belief formation and trading behavior to amplify systemic instability. The central lesson is not simply that AI agents may trade similarly, but that sufficiently similar agents may come to \emph{see} markets similarly---encoding the same features, neglecting the same signals, and creating common blind spots that remain latent until stress exposes them. The formal separation of representation homogeneity from forecast overlap reveals that the integrated distance between agents' predictions may understate the degree of structural fragility, because it averages over states and can mask severe disagreement compression under stress.

The existence of a positive critical representation threshold implies that the transition from resilience to fragility occurs before agents become identical. Residual heterogeneity in readout parameters, risk preferences, and learning rates provides a buffer, but this buffer has finite capacity. Once representation distances fall below the critical level, synchronized forecast errors cascade through order-flow concentration and nonlinear liquidity responses into elevated tail risk.

The experimental framework---combining factorial identification, matched forecast-distance designs, and dynamic convergence simulations---provides multiple complementary lenses through which this mechanism can be tested. Preserving diversity in how market participants represent and process information may therefore be an indispensable stabilizer in AI-dominated financial ecosystems.

\section*{Appendix A: Aligned Representation Distance}

Hidden features in neural representations may be equivalent up to permutation. In such cases, the raw Euclidean distance between feature vectors may exaggerate representation differences even when the encoded information is similar.

To address this issue, we construct an aligned representation distance.

Let \(\mathcal P_K\) denote the set of \(K\times K\) permutation matrices.
We define

\begin{equation}
d_{\mathrm{repr}}^{\mathrm{align}}(i,j)
=
\inf_{\Pi\in\mathcal P_K}
\left(
\int_{\mathcal S}
\|h_i(S)-\Pi h_j(S)\|_2^2
\, d\mu(S)
\right)^{1/2}.
\label{eq:repr_align}
\end{equation}

Operationally, the empirical approximation becomes

\begin{equation}
d_{\mathrm{repr}}^{\mathrm{align}}(i,j)
=
\min_{\Pi\in\mathcal P_K}
\left(
\frac{1}{T}
\sum_{t=1}^{T}
\|h_i(S_t)-\Pi h_j(S_t)\|_2^2
\right)^{1/2}.
\end{equation}

This aligned metric allows hidden units to be reordered before computing distances,
thereby isolating differences in encoded information rather than superficial feature
indexing.

In the robustness analysis, all representation homogeneity statistics are recomputed using
\(d_{\mathrm{repr}}^{\mathrm{align}}\) to verify that the findings are not driven by arbitrary permutations of hidden features.

\bibliographystyle{apalike}
\bibliography{references}

\end{document}